\documentclass{appolx}
\usepackage{epsfig}
\begin{document}
\title{Specific heat of the mixed spin-1/2 and spin-$S$ Ising model 
with a rope ladder structure\thanks{Presented at CSMAG'07 Conference, 
Ko\v{s}ice, 9-12 July 2007}}
\author{J.~KI\v{S}\v{S}OV\'A and J.~STRE\v{C}KA 
\address{Department of Theoretical Physics and Astrophysics, Faculty of Science, \\ 
P. J. \v{S}af\'{a}rik University, Park Angelinum 9, 040 01 Ko\v{s}ice, Slovak Republic}}
\maketitle
\begin{abstract}
The mixed spin-1/2 and spin-$S$ ($S>1/2$) Ising model on a rope ladder is examined by combining 
two exact analytical methods. By the decoration-iteration mapping transformation, this mixed-spin system is firstly transformed to a simple spin-1/2 Ising model on the two-leg ladder, which is 
then exactly solved by the standard transfer-matrix method. The thermal variations 
of the zero-field specific heat are discussed in particular.
\newline
\end{abstract}

\section{Introduction}
One-dimensional Ising models are of fundamental scientific interest partly on behalf of their exact solubility \cite{thom79} and partly due to the fact that they represent the simplest lattice-statistical models, which have found rich applications in seemingly diverse research areas \cite{thom88}. In the present work, we shall provide the exact solution for the mixed spin-1/2 and spin-$S$ ($S>1/2$) Ising model on a rope ladder, which represents a rather novel magnetic structure emerging in several insulating bimetallic coordination compounds \cite{ohba94}.

\section{Model and its exact solution}
\begin{figure}[t]
     \begin{center}
       \includegraphics[width = 0.75\textwidth]{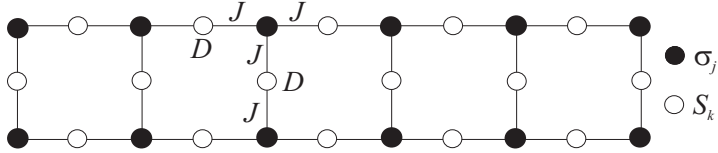} 
       \caption{\small A small fragment of the mixed-spin Ising model on the rope ladder. 
       The full circles denote lattice positions of the vertex spins $\sigma_j = \pm 1$, 
       while the empty ones label lattice positions of the decorating spins 
       $S_k = -S,-S+1, \ldots , +S$.}
       \label{fig1}
     \end{center}
\end{figure}

Consider the spin-1/2 Ising ladder of $2N$ sites with each of its horizontal and vertical 
bonds occupied by the decorating spin $S$ ($S>1/2$), as it is schematically shown in Fig.~\ref{fig1}.
For further convenience, the Hamiltonian of the model under investigation can be written 
as a sum of site Hamiltonians 
\begin{eqnarray}
{\cal H} = \sum_{k=1}^{3N} {\cal H}_{k}, \qquad 
{\cal H}_{k} = - J S_{k} (\sigma_{k1} + \sigma_{k2}) - D S_{k}^{2},
\label{1}
\end{eqnarray}
where each site Hamiltonian ${\cal H}_{k}$ involves all the interaction terms of one decorating 
spin. Above, $\sigma_{k \gamma}=\pm 1$ ($\gamma = 1,2$) denote two vertex spins of the rope ladder 
that are nearest neighbours to the decorating spin $S_{k} = -S,-S+1, \ldots , +S$. The parameter 
$J$ labels pairwise exchange interaction between the nearest-neighbour spin pairs, while the 
parameter $D$ measures a strength of the single-ion anisotropy acting on the decorating spins. 

The crucial step of our procedure lies in the calculation of the partition function, 
which can be firstly partially factorized to the expression
\begin{eqnarray}
{\cal Z} = \sum_{\left\{ \sigma_j \right\}} \prod_{j=1}^{3N} \sum^{+S}_{n=-S}   
\exp \left(\beta D n^2 \right) \cosh \left[\beta J n \left(\sigma_{k1} + \sigma_{k2}
\right) \right], 
\label{2} 
\end{eqnarray}
where $\beta = 1/(k_{\rm B} T)$, $k_{\rm B}$ is Boltzmann's constant, $T$ is the absolute temperature and the symbol $\sum_{\left\{\sigma \right \}}$ denotes the summation over all possible spin configurations of the vertex spins $\sigma_{j}$. Next, it is advisable to employ the generalized decoration-iteration mapping transformation \cite{fish59}
\begin{eqnarray}
\sum^{+S}_{n=-S}  \exp \left(\beta D n^2 \right) \cosh \left[\beta J n \left(\sigma_{k1} + \sigma_{k2}
\right) \right] = A \exp \left(\beta R \sigma_{k1} \sigma_{k2} \right), 
\label{3} 
\end{eqnarray}
which physically corresponds to removing all the interaction parameters associated with the decorating spins and replacing them by a new effective interaction $R$ between the remaining vertex spins. 
Since there are four possible spin states for each pair of vertex spins $\sigma_{k1}$ 
and $\sigma_{k2}$ and only two of them provide independent equations from the transformation 
formula (\ref{3}), the mapping parameters $A$ and $R$ have to meet the following conditions
\begin{eqnarray}
A^2 \! \! &=& \! \! \left(\sum_{n=-S}^{+S} \! \! \exp(\beta Dn^2) \cosh(2 \beta J n) \right)
                          \left(\sum_{n=-S}^{+S} \! \! \exp(\beta Dn^2) \right), \label{4a}  \\ 
\beta R \! \! &=& \! \! \frac{1}{2} \ln \left(\sum_{n=-S}^{+S} \! \! \exp(\beta Dn^2) 
\cosh(2 \beta J n) \right) -  \frac{1}{2} \ln \left(\sum_{n=-S}^{+S} \! \! \exp(\beta Dn^2) 
\right)\! \!.  
\label{4b} 
\end{eqnarray}
By substituting the transformation (\ref{3}) into Eq.~(\ref{2}) one gains the relation
\begin{eqnarray}
{\cal Z} (\beta, J, D) = A^{3N} {\cal Z}_{0} (\beta, R),
\label{6} 
\end{eqnarray}
which connects the partition function ${\cal Z}$ of the mixed-spin Ising model on the rope ladder 
with the partition function ${\cal Z}_{0}$ of the simple spin-1/2 Ising model on two-leg ladder with the effective  nearest-neighbour interaction $R$. It is worthwhile to remark that the partition function of the latter model can be rather easily calculated by applying the transfer-matrix method \cite{thom79}, which yields in the thermodynamic limit the following exact result 
\begin{eqnarray}
{\cal Z}_0 = \left \{\cosh (3 \beta R) + \cosh(\beta R) 
          + \sqrt{\left[\sinh (3 \beta R) - \sinh(\beta R)\right]^2 + 4} \right\}^N\!\!.
\label{7} 
\end{eqnarray}
Bearing this in mind, the exact solution for the partition function of the mixed-spin Ising model 
on the rope ladder is formally completed as it can be simply achieved by a mere substitution of the corresponding partition function (\ref{7}) to the relationship (\ref{6}), whereas both mapping parameters $A$ and $R$ have to be taken from Eqs.~(\ref{4a}) and (\ref{4b}), respectively.
 
\section{Results and discussion}

Now, let us focus on thermal variations of the zero-field specific heat with the aim to 
shed light on how this dependence changes with the single-ion anisotropy. Before discussing 
the most interesting results, it should be mentioned that the zero-field specific 
heat can readily be obtained from the exact result for the partition function (\ref{6}) 
with the help of basic thermodynamical-statistical relations. Even although the final 
expression is too cumbersome to write it down here explicitly, it is worthy to note 
nevertheless that it implies an independence on a sign of the exchange interaction $J$. 
In this respect, the temperature dependences of the zero-field specific heat discussed below 
remain in force regardless of whether ferromagnetic ($J>0$) or ferrimagnetic ($J<0$) 
spin system is considered. 

For illustration, Fig.~\ref{2}a) and b) depict thermal variations of the zero-field specific 
heat for the rope ladder with the decorating spins $S=1$ and $S=3/2$.
\begin{figure}[ht]
     \begin{center}
       \includegraphics[width = 0.52\textwidth]{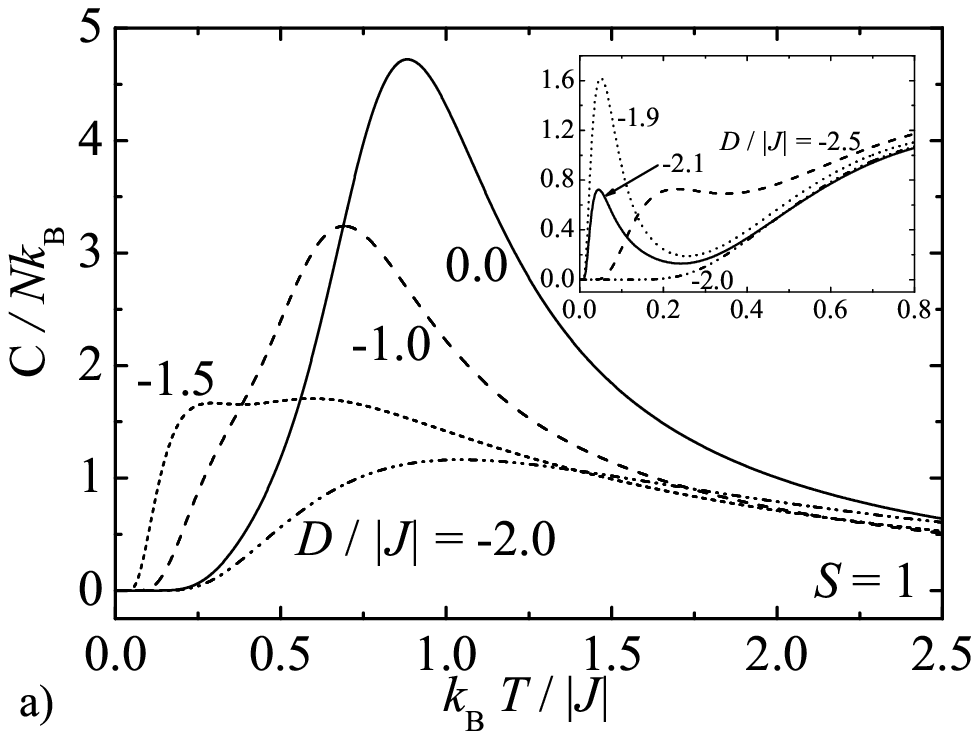} \hspace{-0.8cm}
       \includegraphics[width = 0.52\textwidth]{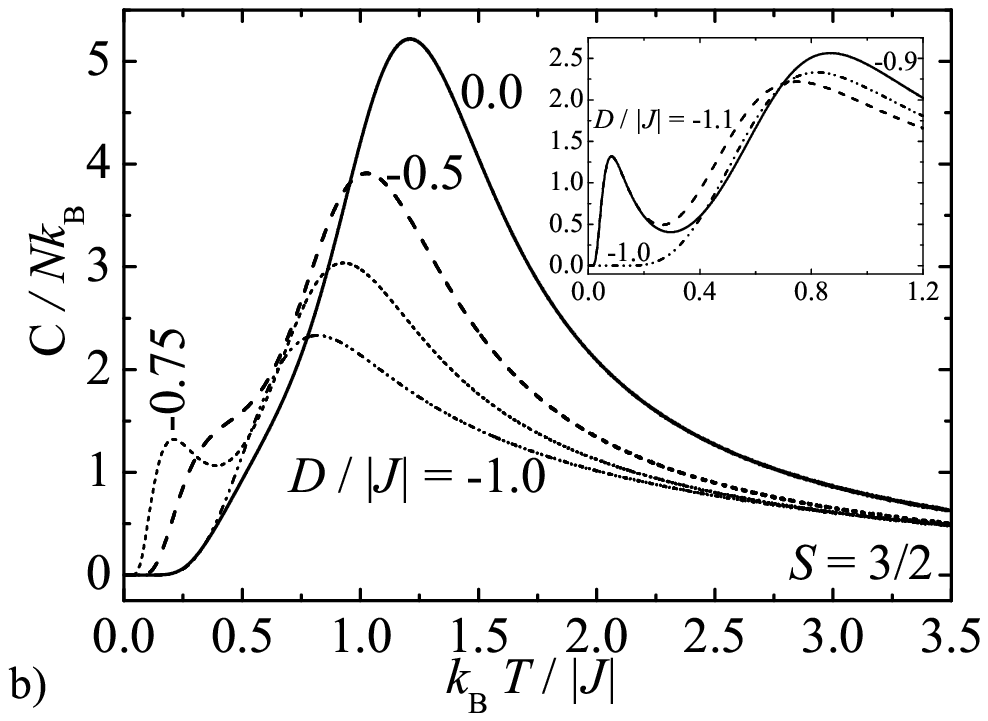}
       \caption{\small The temperature dependences of zero-field specific heat for several
       single-ion anisotropies and two different spin values $S=1$ (Fig.~\ref{fig2}a) and 
       $S=3/2$ (Fig.~\ref{fig2}b).}
       \label{fig2}
     \end{center}
\end{figure}
As one can see, the same general trends can be observed in both figures: the round maximum 
diminishes and shifts towards lower temperatures by decreasing the single-ion anisotropy. However, 
the most interesting temperature dependences can be found at sufficiently negative single-ion anisotropies close to $D/J = -2.0$ for the case with $S=1$ and respectively, $D/J = -1.0$ for 
the case with $S=3/2$. In the vicinity of both these boundary values, the double-peak specific 
heat curves (low-temperature peaks are for clarity shown in the insets) appear due to the competition between two different spin configurations sufficiently close in energy. Actually, it can easily 
be verified that the general condition ensuring an energetic equivalence between two different 
spin states of decorating spins reads $D_{S \leftrightarrow S-1} = -2 J/(2S-1)$. It should be 
also mentioned that the zero-field specific heat curves with a single maximum are recovered 
upon further decrease of the single-ion anisotropy. In conclusion, it seems valuable to extend 
our calculation to more general case with the non-zero magnetic field.

\begin{center}
{\bf Acknowledgments}
\end{center}
This work was supported by the Slovak Research and Development Agency under the contract 
No. LPP-0107-06 and the grant No. VEGA 1/2009/05.

\end{document}